# A Novel Framework for Software Defined Wireless Body Area Network

Khalid Hasan[1], Xin-Wen Wu[1], Kamanashis Biswas[1], Khandakar Ahmed[2]
*School of Information and Communication Technology*[1]
Griffith University, Gold Coast, Australia[1]
*College of Engineering and Science*[2]
Victoria University, Melbourne, Australia[2]
khalid.hasan@griffithuni.edu.au

*Abstract* — **Software Defined Networking (SDN) has gained huge popularity in replacing traditional network by offering flexible and dynamic network management. It has drawn significant attention of the researchers from both academia and industries. Particularly, incorporating SDN in Wireless Body Area Network (WBAN) applications indicates promising benefits in terms of dealing with challenges like traffic management, authentication, energy efficiency etc. while enhancing administrative control. This paper presents a novel framework for Software Defined WBAN (SDWBAN), which brings the concept of SDN technology into WBAN applications. By decoupling the control plane from data plane and having more programmatic control would assist to overcome the current lacking and challenges of WBAN. Therefore, we provide a conceptual framework for SDWBAN with packet flow model and a future direction of research pertaining to SDWBAN.**

*Keywords* — *SDWBAN, IoT, Healthcare, SDN, WBAN*

## I. INTRODUCTION

Tremendous and rapid advancement of internet technologies has been facilitating in numerous ways in daily life. Now-a-days, the Internet is not just about connecting end users, it is also interconnecting physical objects or 'Things' [1]. Development of such technologies has introduced the term "Internet of Things (IoTs)" and many sectors of modern life now implicitly or explicitly are related to IoTs. From building and home automation, smart city technology, industrial automation to healthcare, IoT touches every aspect of daily life. Applications and advancements of IoTs in the healthcare industry introduces us with Wireless Body Area Network (WBAN), which is a significant addition to IoT platform.

WBAN technology involves health-related sensors, which are installed near the patient's body or on the body or even in some cases, implanted inside [2]. These sensors are interconnected via short-range wireless technology such as ZigBee, WiFi, Bluetooth etc. and form a wireless body area network. A mobile or Personal Digital Assistant (PDA) or any other device works as a gateway, which collects data from the sensors and forwards the data to a remote online server where data processing and analysis takes place with the assistance of some medical software applications. The communication link between gateways to online servers could be based on telecommunication networks such as LTE, LTE-A, WLAN, WiMax or Satellite. Despite various development in WBAN, still it poses lots of challenges [3]. Complexity of management, static architecture, dynamic reconfiguration, resource utilization, coexistence and interference handling, efficient traffic management, security and authentication, energy efficiency etc. are the major challenges experienced in WBANs. Failures in dealing with these challenges may lead to poor network performance. Issues and challenges of WBAN, can be overcome by deploying Software Defined Networking (SDN) based WBAN. SDN separates the data and control plane of the network which provides more programmatic control of the network [4]. Having more programmatic control over the network would ease the complexity of network management by incorporating dynamic configurability, traffic priority policy and efficient resource allocations.

In this paper, we propose a new framework for WBANs based on SDN. We derive the motivation for this new framework based on lack of current WBAN architectures and challenges such as heterogeneous WBAN traffic handling, static architecture, mobility management, traffic priority management, secured authentication, network reconfiguration, energy efficiency etc. The rest of the paper is organized as follows: Section II discusses the limitations of current WBAN. Section III provides the discussion on the benefits of using SDN in WBAN and Section IV proposes a conceptual framework of SDN based WBAN and proposes a packet dissemination model. Finally, concluding remarks and future direction of research are provided in section V.

## II. LIMITATIONS OF CURRENT WBAN

Wireless Body Area Network (WBAN) consists of different health related sensors attached to the body or in some cases, sensors are implanted in the body. These sensors are used to monitor different physiological data such as ECG, blood pressure, body temperature and heart rate. In general, WBAN architecture can be classified into three subsystems logically: data collection, transmission and processing. Body sensors are responsible to collect physiological data and send the data to the gateway whereas the gateway transmits data to the remote server for processing. Different sensors based on the application, are equipped with data acquiring functionality, wireless data transmission module, energy supply module and micro-controller to coordinate the periphery module and execute the function of data processing. Although the system looks intuitive, there are still some drawbacks.

The heterogeneous nature of WBAN requires different sensors to be installed for measuring physiological parameters. These sensors could be from multiple vendors. Different vendors employ different sensor platforms with various wireless data transmission modules. Typically, devices from multiple vendors do not interoperate with one another [5]. Therefore, a network consists of multiple vendor specific sensors, is complex to manage and maintenance requires high investment when there is a change in the network.

Application based architecture of WBAN is static in nature. The infrastructure and the application are closely coupled [6]. Any changes required in the application intelligence, would require changes in the sensor platform, gateway and server. Ultimately, a secondary defined physical structure would be required when each application needs to deploy its own sensor platform, gateway and remote server from the scratch. Therefore, introducing a new application is not easy, rather it requires a lengthy deployment process. Hence, it creates a barrier towards potential application innovation.

In healthcare application of WBAN, patients usually have the freedom of mobility while body sensors are attached on their body. In such case, sometimes they are in the vicinity or inside the neighboring WBAN. This mobility causes packet loss and increases error rate [7]. Therefore, proper handover mechanism should be implemented in WBAN.

In traditional WBAN, traffic priority is maintained through different access mechanisms defined by various standards. There is no administrative control on the traffic flow. In WBAN, we have low priority traffic, emergency and on-demand traffic. In cases of multiple emergency events, where multiple sensors would like to transmit data, would create congestion. As a result, the delay caused by the congestion, might result catastrophically. Above all, diverse traffic pattern need to be handled carefully.

Network reconfiguration is an important part in successful network management. Any addition or changes in the setting should not hinder seamlessness of the network. Current architectures of WBAN management, fails to provide reconfiguration capability. A real example is Nagoya City University Hospital [8]. Individual networks were constructed for each clinical department. Consequently, individually optimized network on different layers made the whole network configuration very complex. Any addition of any equipment or changes in settings required stopping the network, which required a lot of time and effort of hospital staff.

Inefficient resource utilization is another pitfall of the current architecture. According to [6], applied control logic which is embedded into hardware, cannot dynamically improve resource utilization. Specifically, when data collection cannot be controlled dynamically, sensors will continue to transmit data to a remote server even if the data are unwanted. This costs unnecessary energy waste and wastage of network bandwidth.

Although advancement of information technology has improved WBAN systems, dealing with sensitive information remains a challenge. Network architecture tends to be complex when it deals with massive data. Eventually, managing a complex network with proper authentication and secure delivery becomes cumbersome. It makes the system vulnerable to unauthorized access, which is referred as cybercrime [9, 10]. Therefore, in terms of data confidentiality, authenticity, integrity and secure management, a strict and scalable security system is still a matter of concern for WBANs [11].

There have been few projects on the development of WBANs. One of the projects was CodeBlue [12]. This project aimed at providing infrastructure for highly scalable dense network, which delivers data in an ad hoc manner. This project was designed to tackle emergency medical events with self-organizing capability. Although, CodeBlue supports scalable, timeliness and security, the most serious lack is the reliable communication.

Advanced Health and Disaster Aid Network (AID-N) [13] was introduced for health monitoring in a mass casualty incident in real time communication. This architecture introduced three layers of which, the first layer employs an ad-hoc network for collecting vital parameters, the second layer consists of personal servers such as PDA or laptops and a central server is in third layer. One of the prime goals of this project was to be able to efficiently collect data, distribute information on vital signs and locate patients in an extremely fault tolerant manner. However, the system was only able to locate the patients who were near to the base station. Indoor patients were not successfully located and therefore, reliable communication links could not be established. Furthermore, AID-N failed to deal with body area interference due to the wandering behavior of the patients. Thus, it was unable to efficiently route data with this type of information.

Another remote healthcare monitoring project is CareNet [14]. CareNet, aimed to provide reliable and privacy preserving information between patients' home and the health care provider. This project used a heterogeneous platform where data collection, transmission and processing take place based on two-tier wireless network. Wearable body sensors use IEEE 802.15.4 to send data to a backbone router which then forwards data using IEEE 802.11 to healthcare gateway through multi-hop communication. Although, CareNet supports WBAN communication with reasonable reliability, scalability and security, CareNet neglects the issue of realtime communication in emergency health monitoring applications.

For Ambulant patient monitoring, MobiHealth [15] system was designed which uses cellular network. The physiological signs are collected via Bluetooth and

ZigBee. This architecture supports intra-WBAN and beyond WBAN communication, however Mobihealth fails to deal with security and data privacy issues.

### III. BENEFITS OF SDN ON WBAN

This section presents the benefits of SDN on different aspects.

**Getting Rid of Complex Management-** SDN provides more administrative control by separating the control plane from the data plane. In the case of SDN implementation in WBAN, health service providers will have more flexibility and programmatic control over the network. It would create a simplified platform for apps and programs. Centrally control operations allow health service providers to monitor multiple sites from one location. Therefore, individual network team administrators are not required for every site.

**Independence of Vendor-** Open standard based implementation of SDN simplifies the network design and operations. This is because the instructions and commands are provided from the controller instead of vendor specific instructions and protocols [16].

**Data Prioritization-** More administrative controls aid to deal with heterogeneous traffic pattern. Priority can be given based on the situation. Life critical data can be given high priority over the normal monitoring data. Therefore, from and infrastructural standpoint, SDN could be used to prioritize various types of data in WBAN. The SDN controller is capable of reserving bandwidth for delay sensitive applications to ensure guaranteed transmission [6].

**Patient Monitoring-** Patient monitoring system could be more flexible and reliable with the help of SDN. In WBAN, free mobility of the patient requires monitoring as well, as an example, people with dementia disease who have wandering behavior. In [17], a secure monitoring technique through tracking the location of patients' and alarm raising system have been discussed. Therefore, whenever an endpoint joins the network, SDN controller can identify the patient [18]. Patient monitoring controller can locate the endpoint and links the access interface to the virtual network.

**Security-** Cybercrime is another serious threat to healthcare applications. Patients' data are confidential and private; therefore, it demands secured and authorized access only. SDN can provide identification and authentication services to defend against several cyber-attacks in healthcare. For instance, Kanazawa Hospital [19], successfully deployed SDN to create four different virtual network on an individual physical network for the purpose of security. A secured SDN based WBAN architectural system has been proposed in [11] for efficient and secure data delivery.

**Mobility and Accuracy of Location Tracking-** Patients have the freedom of mobility when the body sensors are attached. During mobility, patients sometimes enter the neighboring WBAN. Therefore, accurate tracking of patients' position is important to support mobility. With SDN and a centralized routing algorithm, it is possible to get the accurate location information [20]. Through a mobility management protocol, centralized controller is capable of updating networking policies on the fly. Consequently, controllers update the flow tables with updated routing decision to ensure optimal network performance [21].

**Improvement in Energy Efficiency-** Energy efficiency is an important part to deal with in any wireless sensor network deployment and when it comes to healthcare, it demands significant attention. SDN controller can determine the best routing policy for the sensors. In addition, when a sensor node is about to run out of battery, it can inform the controller to make changes in the routing table [22]. With the help of SDN controller, traffic management, resource allocation and Quality of Service (QoS) can be achieved efficiently with a lower energy overhead [21]. Since SDN can provide an overall view of the network, controller can dynamically activate and deactivate the sensors, customize the policies to satisfy application demands and meanwhile reduce the energy consumption [6].

### IV. PROPOSED FRAMEWORK

This section describes and presents the logical architecture of SDN for WBAN as well as a conceptual framework of SDWBAN.

#### A. Logical Description of SDN for WBAN

In general, SDN has three planes logically, such as data plane, control plane and application plane. Fig.1 presents logical architecture of SDN for WBAN followed by detail description.

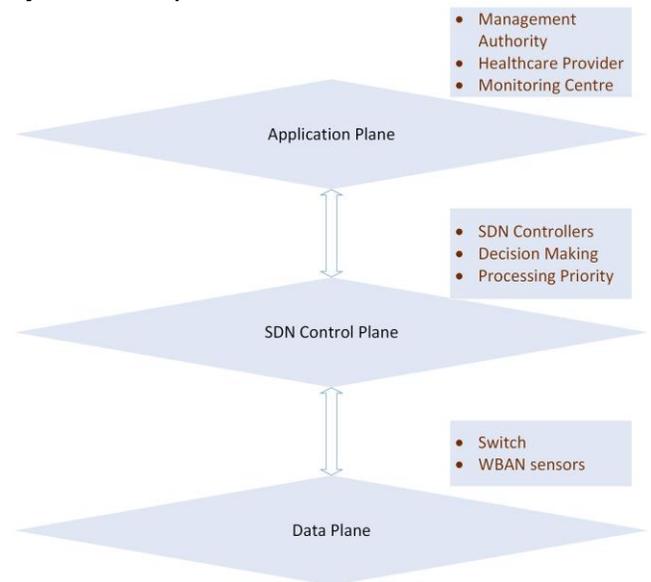

Fig.1. Logical Architecture of SDN for WBAN

**Data Plane-** Data plane of the SDWBAN architecture consists of sensors and switches. Sensors send data to switches and each switch is associated with local controller to retrieve packet forwarding instructions. Packet dissemination control information is directed with a set of rules called flow commands provided by the SDN controller. Upon receiving forwarding commands, switches send data packets to appropriate destinations. Every single switch is responsible for a patient which makes the authentication task simple. Communication between switch and sensors could be based on short range communication technologies such as IEEE 802.15.6, IEEE 802.15.4, Bluetooth etc. Controller can install data prioritizing policy for switches to timely deliver emergency data.

**Control Plane-** The controller controls the whole network management activities. It can capture an overall picture of the network. When the controller receives a request from switch with packet forwarding instructions, controller has the capability to classify the packet based on application and priority. Based on the instructions received from the controller, the switch updates its flow table. Due to scalable nature and heterogeneity of WBAN, whenever a new patient is admitted in an elderly home, the controller can manage packet forwarding instructions regardless of vendor specific devices. It can also translate the new entity so that the network management systems are not hampered.

**Application Plane-** This plane belongs to the management authority. From this plane, health care providers can monitor their patients. They can also define, install, operate and manage new WBAN application sets.

*B. Conceptual Framework of SDWBAN*

To implement Software Defined Networking (SDN) in Wireless Body Area Network (WBAN) we can consider the following scenario as depicted in Fig. 2.

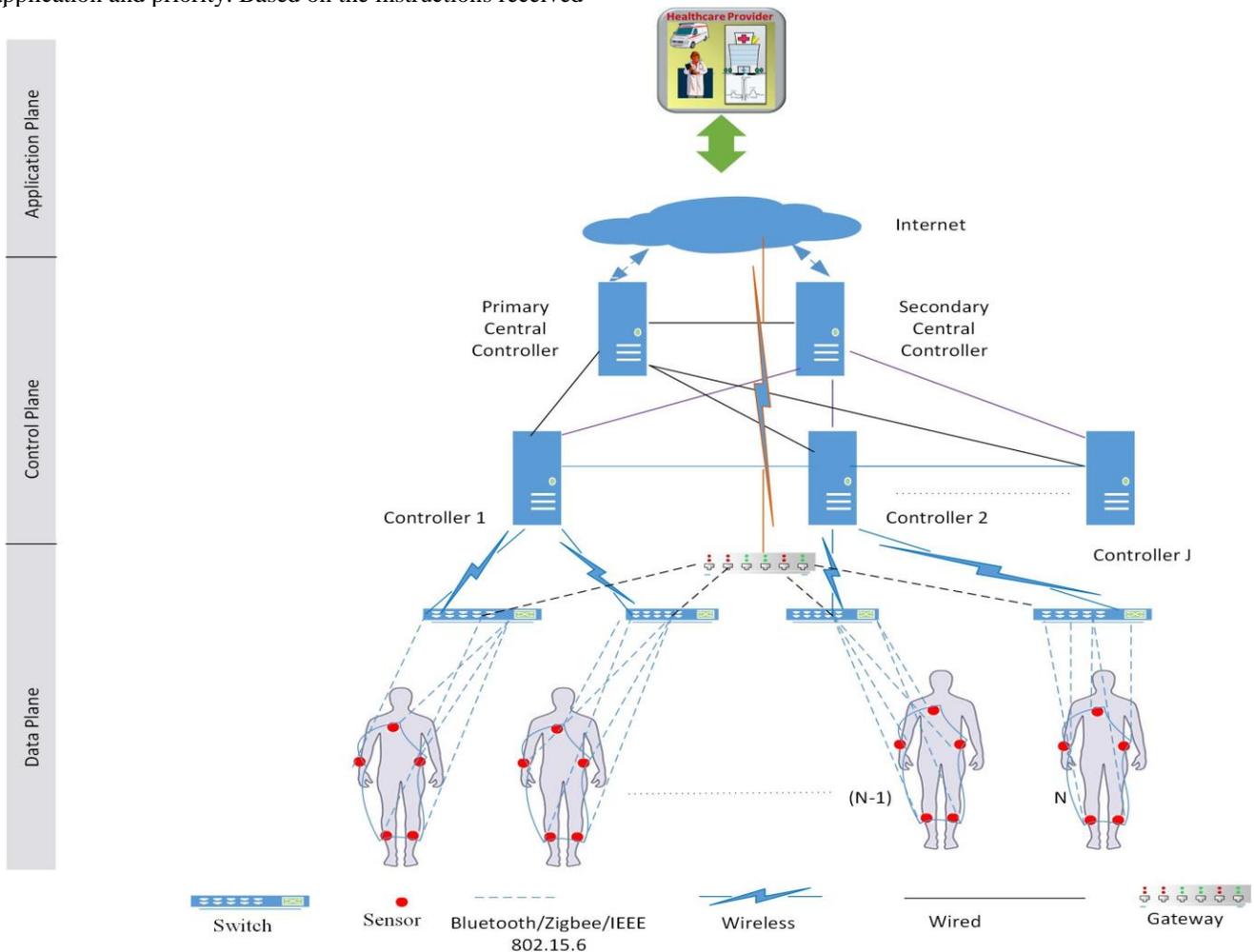

Fig.2. Conceptual Architecture of SDWBAN

Different WBAN sensors are attached on or implanted in a human body. Let us say, there are *N* number of patients in an elderly home. Sensors are responsible to collect and forward data to a switch. In this case, a switch is another sensor node where SDN functionalities are implemented to communicate with the controller. We will address the switches as SDN enabled switch. Each SDN enabled switch associates itself with a Local Controller (LC) through OpenFlow protocol based wireless link. LCs can support multiple SDN enabled switches. In the above scenario, we have *J* number of LCs where *J* is much smaller than the number of SDN enabled switch. LCs are inter-connected (wired connection) so that, in case of the failure of one controller, switches can associate themselves with the next available controller to retrieve control information. In addition, to avoid single point of failure, a secondary central controller could also serve the local controllers in absence of a primary controller. Based on flow information received from the controller, SDN enabled switch then forwards the data to a gateway. The gateway connects to a cloud using long range communication technologies such as LTE, LTE-A, WiMax etc. In this case, we can assume each patient with multiple sensor nodes forms a cluster and a SDN enabled switch acts as the cluster head. In an elderly home, there are many patients and sensors of various applications installed on or inside their body. One patient body could have multiple application sensors such as heart rate, temperature, glucometer, blood pressure etc. These sensors send data to SDN enabled switches at different time intervals. Switches forward the data packet to a gateway and aggregated data from the gateway goes to a cloud. Medical service providers access to the cloud and monitor their concerned patients.

*C. Packet Dissemination Model*

Based on priority, WBAN traffic can be classified into two categories: normal traffic and emergency traffic. Normal traffic are periodic monitoring data which provides general physiological data in periodic manner, whereas emergency traffic are generated when physiological condition exceeds a predefined threshold. Emergency data traffic needs to be addressed in the first most priority basis to avoid undesired circumstances. Upon receiving data packets from WBAN sensors, SDN enabled switch can classify the data packet as normal or emergency based on predefined physiological threshold. In case of normal data packet, table lookup task is performed for flow matching. If the received data packet flow is matched, corresponding switch initiates an action based on the flow command. On the other hand, if there is a table miss according to pre-installed flow table, switches can request for instructions from the controller for the corresponding packet. The SDN controller then creates a new flow command and the switches update the flow table and take appropriate actions. However, when an emergency data packet is identified, SDN enabled switch finds a definite flow match and therefore takes the prompt action. The controller broadcasts command to all SDN enabled switches regarding emergency data so that, emergency data packet gets priority over the normal data traffic. Fig. 3 explains data packet exchange mechanism.

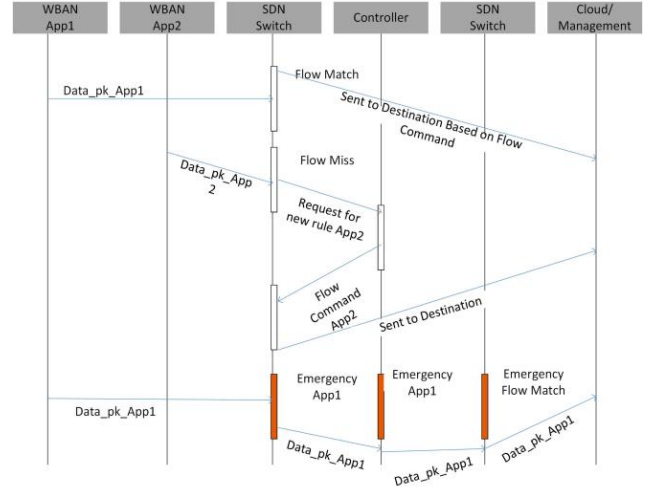

Fig. 3. Packet dissemination model for SDWBAN

For instance, data_pk_App1 from WBAN App1 is received at the switch and the switch identifies the data packet as normal data. Flow table lookup task is performed and upon finding a match in the flow table it forwards data packet to the cloud where healthcare provider can have access. When second data packet arrives at the switch form WBAN App2, it again performs flow table lookup task and finds a flow miss. The switch then initiates a packet-in requests for new flow rule for data_pk_App2. Upon receiving a new command from the corresponding controller, SDN enabled switch takes the necessary action and updates the flow table. However, in case of emergency data, data_pk_App1, all switches are immediately informed regarding this emergency data packet. Upon finding a definite flow match, corresponding switch take prompt action.

## V. CONCLUSION & FUTURE WORK

In this paper, we have presented a framework for SDN based WBAN upon highlighting the necessity of incorporating SDN into WBAN. We have explained the logical and conceptual architecture from the perspective of SDN functionalities. We have proposed a model to handle normal and emergency data packet. This paper also contributes to identify the possible improvement sectors for WBAN with the help of SDN. For better network management with differentiated traffic pattern, security and authentication, energy efficiency, mobility; SDN is an appealing choice.

The future works will validate the feasibility of the proposed framework in terms of physiological requirements of different WBAN applications. It is also important to find optimal number of controllers and switches for WBAN regarding maintain the Quality of Service. Future works could also include a test-bed implementation.

REFERENCES

[1] S. Tyagi, A. Agarwal, and P. Maheshwari, "A conceptual framework for IoT-based healthcare system using cloud computing," in *Cloud System and Big Data Engineering (Confluence), 2016 6th International Conference*, 2016, pp. 503-507: IEEE.
[2] A. Darwish and A. E. Hassanien, "Wearable and implantable wireless sensor network solutions for healthcare monitoring," *Sensors,* vol. 11, no. 6, pp. 5561-5595, 2011.
[3] C. A. Chin, G. V. Crosby, T. Ghosh, and R. Murimi, "Advances and challenges of wireless body area networks for healthcare applications," in *Computing, Networking and Communications (ICNC), 2012 International Conference on*, 2012, pp. 99-103: IEEE.
[4] M. Uddin, *Toward open and programmable wireless network edge*. Old Dominion University, 2016.
[5] A. Drescher, "A survey of software-defined wireless networks," *Dept. Comput. Sci. Eng., Washington Univ. St. Louis, St. Louis, MO, USA, Tech. Rep,* pp. 1-15, 2014.
[6] L. Hu, M. Qiu, J. Song, M. S. Hossain, and A. Ghoneim, "Software defined healthcare networks," *IEEE Wireless Communications,* vol. 22, no. 6, pp. 67-75, 2015.
[7] G. Cova, H. Xiong, Q. Gao, E. Guerrero, R. Ricardo, and J. Estevez, "A perspective of state-of-the-art wireless technologies for e-health applications," in *IT in Medicine & Education, 2009. ITIME'09. IEEE International Symposium on*, 2009, vol. 1, pp. 76-81: IEEE.
[8] NEC. (2018, February 9). *Software-Defined Networking (SDN) Solution Nagoa City University Hospital*. Available: http://au.nec.com/en_AU/media/docs/case-studies/nec-sdn-case-study-nagoyai-city-university-hospital.pdf
[9] G. B. Satrya, N. D. Cahyani, and R. F. Andreta, "The detection of 8 type malware botnet using hybrid malware analysis in executable file windows operating systems," in *Proceedings of the 17th International Conference on Electronic Commerce 2015*, 2015, p. 5: ACM.
[10] G. B. Satrya and S. Y. Shin, "Optimizing rule on open source firewall using content and pcre combination," *Journal of Advances in Computer Networks,* vol. 3, no. 3, pp. 308-314, 2015.
[11] M. Al Shayokh, A. Abeshu, G. Satrya, and M. Nugroho, "Efficient and secure data delivery in software defined WBAN for virtual hospital," in *Control, Electronics, Renewable Energy and Communications (ICCEREC), 2016 International Conference on*, 2016, pp. 12-16: IEEE.
[12] V. Shnayder, B.-r. Chen, K. Lorincz, T. R. Fulford-Jones, and M. Welsh, "Sensor networks for medical care," 2005.
[13] T. Gao *et al.*, "The advanced health and disaster aid network: A light-weight wireless medical system for triage," *IEEE Transactions on biomedical circuits and systems,* vol. 1, no. 3, pp. 203-216, 2007.
[14] S. Jiang *et al.*, "CareNet: an integrated wireless sensor networking environment for remote healthcare," in *Proceedings of the ICST 3rd international conference on Body area networks*, 2008, p. 9: ICST (Institute for Computer Sciences, Social-Informatics and Telecommunications Engineering).
[15] K. Wac *et al.*, "Mobile patient monitoring: the MobiHealth system," in *Engineering in Medicine and Biology Society, 2009. EMBC 2009. Annual International Conference of the IEEE*, 2009, pp. 1238-1241: IEEE.
[16] H. Infrastructure. ( 2018, February 22). *Benefits of Software-Defined Networking in Healthcare*. Available: https://hitinfrastructure.com/features/benefits-of-software-defined-networking-in-healthcare
[17] V. Varadharajan, U. Tupakula, and K. Karmakar, "Secure Monitoring of Patients with Wandering Behaviour in Hospital Environments," *IEEE Access,* 2017.
[18] T. Slattery. (February 7). *Healthcare & SDN : A Good Match?* Available: https://www.nojitter.com/post/240171088/healthcare--sdn-a-good-match
[19] N. Cranford. (February 17). *How SDN can benefit healthcare*. Available: https://www.rcrwireless.com/20170714/software/how-sdn-can-benefit-healthcare-tag99
[20] B. T. De Oliveira, L. B. Gabriel, and C. B. Margi, "TinySDN: Enabling multiple controllers for software-defined wireless sensor networks," *IEEE Latin America Transactions,* vol. 13, no. 11, pp. 3690-3696, 2015.
[21] M. A. Hassan, Q.-T. Vien, and M. Aiash, "Software defined networking for wireless sensor networks: a survey," *Advances in Wireless Communications and Networks,* vol. 3, no. 2, pp. 10-22, 2017.
[22] Y. Choi, Y. Choi, and Y.-G. Hong, "Study on coupling of software-defined networking and wireless sensor networks," in *Ubiquitous and Future Networks (ICUFN), 2016 Eighth International Conference on*, 2016, pp. 900-902: IEEE.